\documentclass[%
reprint,
amsmath,amssymb,
aps,
prx
]{revtex4-2}

\usepackage{graphicx}
\usepackage{dcolumn}
\usepackage{bm}

\usepackage[colorlinks,linkcolor=blue,anchorcolor=blue,citecolor=blue]{hyperref}
\usepackage{graphicx}
\usepackage{dcolumn}
\usepackage{bm}
\usepackage{textgreek}
\usepackage{braket}
\usepackage{dsfont}
\usepackage{array}
\usepackage{booktabs}
\usepackage{rotating}
\usepackage{gensymb}
\usepackage[colorinlistoftodos,prependcaption,textsize=tiny]{todonotes}
\usepackage{nicefrac}
\usepackage{mathtools}
\usepackage{hyperref}
\usepackage{bbm}

\begin{document}

\preprint{APS/123-QED}

\title{High-rate discrete-modulated continuous-variable quantum key distribution with composable security
}

\author{Mingze Wu$^{1,\ddag}$}
\author{Yan Pan$^{2, \ddag}$}
\author{Junhui Li$^{1}$}
\author{Heng Wang$^{2}$}
\author{Lu Fan$^{1}$}
\author{Yun Shao$^{2}$}
\author{Yang Li$^{2}$}
\author{Wei Huang$^{2}$}
\author{Song Yu$^{1}$}
\author{Bingjie Xu$^{2}$}
\email{Correspondence: xbjpku@163.com}
\author{Yichen Zhang$^{1,}$}%
\email{Correspondence: zhangyc@bupt.edu.cn}

\affiliation {$^{1}$State Key Laboratory of Information Photonics and Optical Communications, School of Electronic Engineering, Beijing University of Posts and Telecommunications, Beijing 100876, China\\
$^{2}$National Key Laboratory of Security Communication, Institute of Southwestern Communication, Chengdu 610041, China
}
\affiliation{$^\ddag$These authors contribute equally to this work.}




\date{\today}

\begin{abstract}

Continuous-variable quantum key distribution holds the potential to generate high secret key rates, making it a prime candidate for high-rate metropolitan quantum network applications. However, despite these promising opportunities, the realization of high-rate continuous-variable quantum key distribution systems with composable security remains an elusive goal. Here, we report a discrete-modulated continuous-variable quantum key distribution system with a composable secret key rate of 18.93 Mbps against collective attacks over a 25 km fiber channel. This record breaking rate is achieved through the probability shaped 16QAM-modulated protocol, which employs semidefinite programming to ensure its composable security. Furthermore, we have employed a fully digital and precise quantum signal processing technique to reduce excess noise to extremely low levels, thereby facilitating efficient broadband system operation. While ensuring low complexity and cost, our system achieves a performance advantage of over an order of magnitude compared to previous continuous-variable quantum key distribution systems, providing a promising solution for future deployment of quantum key distribution.

\end{abstract}


\maketitle




\section{Introduction}
Quantum Key Distribution (QKD) is a pioneering method for secret key distribution with information-theoretical security between two remote parties\ \cite{1984Quantum,pirandola2020advances,xu2020secure,portmann2022security}.
Continuous-variable (CV) QKD is a promising technological pathway, which can achieve high key rates within metropolitan areas and exhibits strong compatibility with optical communication systems\ \cite{zhang2024continuous,usenko2025continuous}.  CV-QKD can adopt two typical modulation formats: Gaussian-modulation\ \cite{ralph1999continuous,grosshans2002continuous,weedbrook2004quantum} and discrete-modulation\ \cite{leverrier2009,li2018user,leverrier2019,lin2019,denys2021explicit}. Gaussian-modulated CV-QKD protocols have advanced significantly in terms of theoretical security analysis\ \cite{leverrier2015composable,leverrier2017security,pirandola2021composable,pirandola2024improved}, and various high-performance systems implementing these protocols have been reported \cite{jouguet2013experimental,zhang2019integrated,zhang2020long,jain2022practical,tian2022experimental,pi2023,hajomer2024long,bian2024Continuous}.


Despite these advancements, development of a high-rate experimental CV-QKD system with composable security remains a challenge. Gaussian-modulated protocol requires thousands of constellations to approximate a continuous Gaussian distribution\ \cite{jouguet2012}, which demands sophisticated modulation devices and rigorous classical error correction programs\ \cite{leverrier2009}. These intricate processes present a significant challenge in suppressing excess noise, particularly at high repetition rates. In contrast, discrete-modulated CV-QKD  protocols employ a smaller constellation space, thereby enhancing their compatibility with high-speed wireline components \cite{hajomer2024continuous2}. This advantage allows discrete-modulated protocols to achieve higher repetition frequencies and lower excess noise. More importantly, it compensates for the performance degradation introduced by composable security, making discrete-modulated protocols a strong candidate for addressing the demands of high-rate CV-QKD systems with composable security.

Over the past few years, significant progress has been made in proving the asymptotic security of discrete-modulated CV-QKD using various methods\ \cite{li2018user,leverrier2019,lin2019,denys2021explicit}. Building on these theoretical foundations, various experimental implementations of discrete-modulated CV-QKD with asymptotic security have been reported\ \cite{wang2022sub,pan2022experimental,pereira2022probabilistic,tian2023high,roumestan2024shaped}. Recently, the composable security of quadrature phase shift keying (QPSK) modulated CV-QKD protocols has also been proven\ \cite{kanitschar2023finite,bauml2024security}, marking another important milestone in enhancing the security of these systems. These theoretical advancements make it possible to implement discrete-modulated CV-QKD systems with composable security. However, QPSK-modulated protocol using small constellations, falls short in meeting the performance requirements of high-rate systems. Therefore, there is a pressing need to address this challenge and explore alternative modulation techniques or system architectures that can support high-rate CV-QKD with composable security.


\begin{figure*}[t]
	\centering
	\includegraphics[width=18cm]{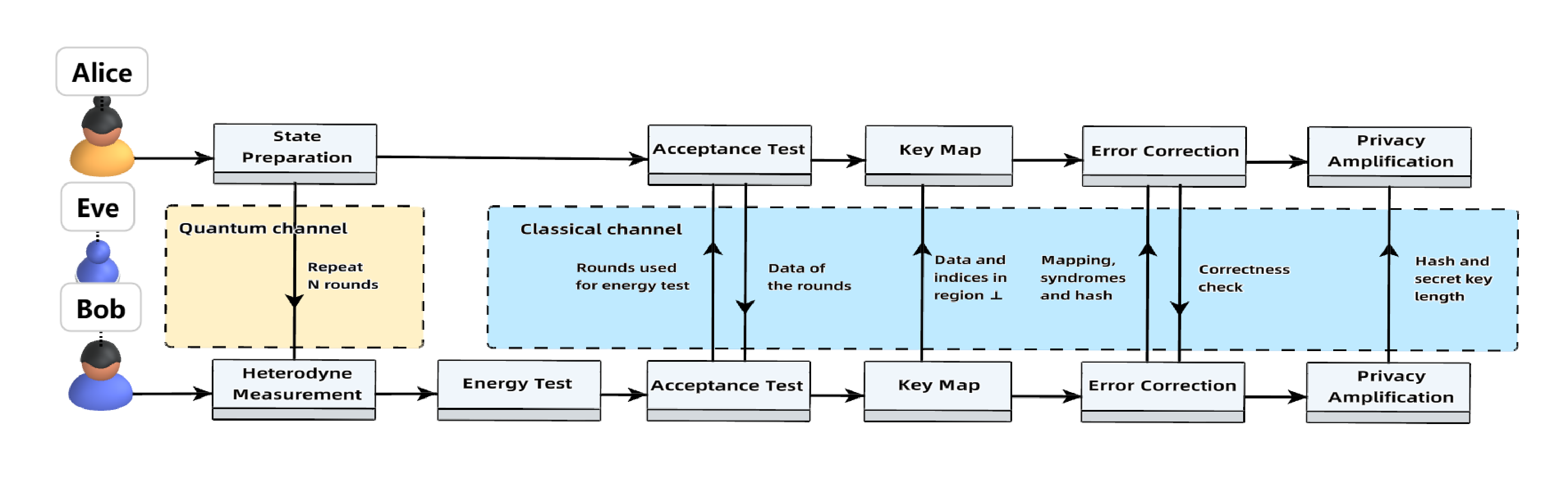}
	\caption{\label{lct}
		Composable discrete-modulated CV-QKD protocol. The protocol mainly include state preparation, heterodyne measurement, energy test, acceptance test, key map, error correction, and privacy amplification, specified in Sec.\ \ref{set2} A.
	}
\end{figure*}


In this paper, we address the critical challenge of enhancing the key rate in discrete-modulated CV-QKD systems with composable security. Recognizing that approximate Gaussian distribution outperforms uniform distribution in the quantum state preparation process of CV-QKD, we adopt probability-shaped 16 quadrature amplitude modulation (QAM) as a strategy to break through the performance limitation of discrete modulated CV-QKD system. Given the lack of a security analysis for 16QAM-modulated CV-QKD protocols with composable security, we embark on a comprehensive theoretical analysis using an advanced security analysis method grounded in semidefinite programming (SDP)\ \cite{kanitschar2023finite}. Building upon this theory, we proceed to conduct an experimental demonstration of the 16QAM-modulated CV-QKD protocol. To enhance the system's flexibility and improve its excess noise suppression performance, a fully digital and high-precision quantum signal processing method is proposed and successfully validated. Notably, our system achieves a remarkable composable key rate of 18.93 Mbps over a 25 km fiber channel. This performance surpasses previous CV-QKD systems by more than an order of magnitude, and is competitive with the most advanced high-rate discrete-variable QKD systems. Our findings not only demonstrate the feasibility of high key rates CV-QKD systems with composable security but also pave the way for promising future deployments of QKD technologies.

This paper is structured as follow. In Sec. \ref{set2}, CV-QKD protocol with 16QAM modulation is introduced and its composable security is analyzed. Modeling the channel, protocol performance is simulated. In Sec. \ref{set4}, experimental system for the protocol is demonstrated. Lastly, discussions are offered and the work is concluded in Sec. \ref{set5}.

\begin{figure*}[t]
	\centering
	\includegraphics[width=16cm]{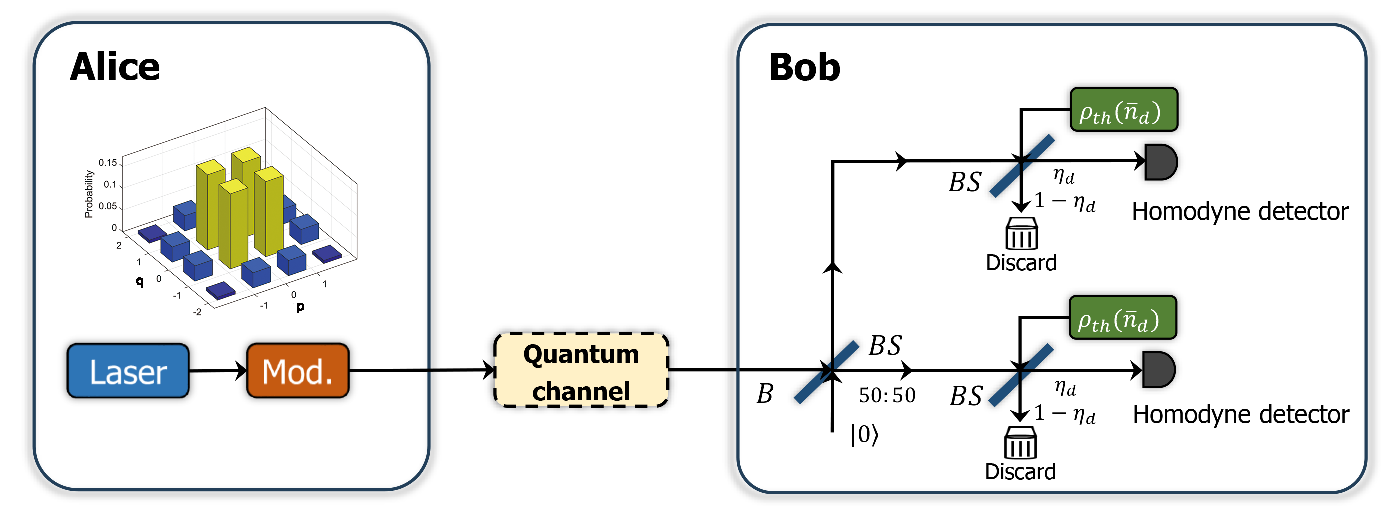}
	\caption{\label{EB}
		Schematic diagram of 16QAM-modulated CV-QKD protocol with trusted detector model, where $\rho_{th}(\bar{n}_d)$ is a thermal state with average photon number $\bar{n}_s$. Mod., modulator; BS, beam splitter; $\eta_{d}$, detection efficiency. The inset of Alice shows the probability distribution of the modulation format.
	}
\end{figure*}
\section{\label{set2}Probability-shaped 16QAM-modulated CV-QKD protocol with composable security}
In this section, probability-shaped 16QAM-modulated CV-QKD protocol with composable security is described. Following this, analysis of its theoretical security is presented. To evaluate performance of the composable 16QAM-modulated CV-QKD protocol, the channel is theoretically modeled as a noisy and lossy Gaussian channel, simulating protocol performance.

\subsection{Protocol Description}
Firstly, composable 16QAM-modulated CV-QKD design is shown in Fig.\ \ref{lct}. The protocol is described as follows:

$(1)$ State preparation. For each round, Alice prepares one of the sixteen coherent states $\left|\alpha_{k}\right\rangle$ which are centered at possible equidistant points with probability $P_{k}$ and transmits it to Bob through the quantum channel, where 
\begin{equation}\label{eq:1}
	\alpha_{k}=q_{k}+i p_{k}, \ \ k\in\{0,1,2,\ldots,15\},
\end{equation}
\begin{equation}\label{eq:1}
	P_{k}={\frac{\exp\Bigl(-\nu(q_{k}^{2}+p_{k}^{2})\Bigr)}{\sum_{k=0}^{15}\exp\Bigl(-\nu(q_{k}^{2}+p_{k}^{2})\Bigr)}}.
\end{equation}
Among them, adjacent states are equally spaced, and we can fix the position of each state by controlling the total variance of each coordinate. $\nu$ needs to satisfy $\nu\textgreater0$.

$(2)$ Heterodyne measurement. After receiving the quantum states, Bob performs trusted heterodyne detection with efficiency $\eta_{d}$, electrical noise $\nu_{el}$ and finite detection range $M$ to obtain measurement result $Y_{j}\in\mathbb{C}$ for each round. The schematic diagram of 16QAM-modulated CV-QKD protocol with trusted detector model is shown in Fig.\ \ref{EB}.

After repeating the above physical steps $N$ times, Alice and Bob perform the classical post-processing steps:

$(3)$ Energy test. Bob performs energy test using $m\ll N$ rounds of raw measurement results. Bob selects test parameter $0<\beta_{test}\leq M$ and number of rounds $l_T$ that may not satisfy the testing condition. If ${\mathrm{Pr}}\lbrack\vert\{Y_{j}\colon \left|Y_{j}\right|^2<\beta_{t e s t}\}\vert\leq l_{T}\rbrack\leq{\epsilon_{ET}}$, which means that most of the weights of transmitted signals are located in a finite dimensional Hilbert space\ \cite{kanitschar2023finite}, the test passes, except for a small error probability $\epsilon_{ET}$. Otherwise, Alice and Bob abort the protocol.

$(4)$ Acceptance test. If the energy test passed, Bob discloses which rounds are used for energy test through classical channel, and Alice discloses the data sent in these rounds to estimate the statistics of their observations. Afterwards, Bob define an acceptance set $S^{AT}$ that can be considered as a list of accepted observations. If the statistical estimators are within the acceptance set, the test passes, except for a small error probability $\epsilon_{AT}$. Otherwise, Alice and Bob abort the protocol.

$(5)$ Key map. For the remaining $n:=N-m$ rounds, Bob performs reverse coordination key map to determine the raw key data $Z$. For this purpose, Bob's measurement results $Y_{j}$ are discretized into a set $z\in\left\{0,1,2,..., 15,\perp\right\}$, discarding the symbols mapped to $\perp$, 

\begin{equation}\label{eq:1}
	\begin{aligned} 
		Z_{j}=
		\begin{cases}
			\vspace{2pt}
			\hspace{0.5em} z &if\ \ Y_{j}\in A_{z}\\
			\vspace{2pt}
			\hspace{0.5em} \perp \ \ &otherwise,
		\end{cases}
	\end{aligned}
\end{equation}
where $A_{z}$ represents the regions illustrated in Fig.\ \ref{xzt25}, and are further elaborated in Appendix\ \ref{set11}.

$(6)$ Error correction. Alice maps the data to $\left\{0,1,2,..., 15\right\}$ according to the corresponding rules
\begin{equation}\label{eq:1}
	x_{j}=k, \ \ if \left|\psi_{j}\right\rangle=\left|\alpha_{k}\right\rangle, 		
\end{equation}
where $k\in\left\{0,1,2,..., 15\right\}$. Alice and Bob publicly communicate over classical channel to reconcile their raw keys $X$ and $Z$. After the error correction, Alice and Bob share the raw key, except for a small portion $\epsilon_{EC}$.

$(7)$ Privacy amplification. Alice and Bob apply two universal hash functions to their raw keys. Except for a small failure probability of $\epsilon_{PA}$, Alice and Bob share the secret key.

\subsection{Security analysis}
Next, we analysis the security of the 16QAM-modulated CV-QKD protocol. Based on the security analysis framework\ \cite{kanitschar2023finite}, secret key length $\ell$ of the 16QAM-modulated CV-QKD protocol with $\epsilon=\epsilon_{\mathrm{EC}}+\mathrm{max}\left\{{\frac{1}{2}}\epsilon_{\mathrm{PA}}+\bar{\epsilon},\epsilon_{\mathrm{ET}}+\epsilon_{\mathrm{AT}}\right\}$-security satisfies


\begin{equation}\label{eq:5}
	\begin{aligned}
		\frac{\ell}{N}\leq\frac{n}{N}&\left[\operatorname*{min}_{\bar{\rho}\in S^{\mathrm{E\&A}}}H(X|E^{\prime})_{\bar{\rho}}-\Delta(w)-\delta(\bar{\epsilon})\right] \\
		&-\delta_{\mathrm{leak}}^{\mathrm{EC}}-\frac{2}{N}\log_{2}\left(\frac{1}{\epsilon_{\mathrm{PA}}}\right),
	\end{aligned}
\end{equation}
where $n$ is the rounds used to generate secret key, $N$ is the total number of rounds, $\delta_{\mathrm{leak}}^{\mathrm{EC}}$ takes the classical error correction cost into account, $\delta(\bar{\epsilon})=2\log_{2}(\mathrm{rank}(\rho_{A})+3)\sqrt{\log_{2}(2/\bar{\epsilon})/n}$, $\bar{\epsilon}$ is the security parameter for smoothing, and $\Delta(w)$ is the determine correction term as
\begin{equation}\label{eq:1}
	\Delta(w):=~\sqrt{w}\,\mathrm{log}_{2}(\vert Z\vert)\,+\,(1\,+\,\sqrt{w})h\left(\frac{\sqrt{w}}{1+\sqrt{w}}\right),
\end{equation}
where $\vert Z\vert$ represents the dimension of key map, $h(\cdot)$ is binary entropy, and for 16QAM-modulated CV-QKD, the bound weight $w$ satisfies
\begin{equation}\label{opt}
	w=\sum_{k=0}^{15}P_{k}\frac{\langle{\hat{n}}_{\beta_{k}}^{2}\rangle-\langle{\hat{n}}_{\beta_{k}}\rangle}{N_{c}(N_{c}+1)},
\end{equation}
where $N_{c}$ is the subspace dimension parameter. The minimization term $\operatorname*{min}_{\bar{\rho}\in S^{\mathrm{E\&A}}}H(X|E^{\prime})_{\bar{\rho}}$ need to be calculated using SDP. $S^{\mathrm{E\&A}}$ contains all states that pass both energy test and acceptance test except for probability $(\epsilon_{ET}+\epsilon_{AT})$. 

To cope with practical experiments, we consider the trusted detector noise model\ \cite{lin2020trusted}, as shown in Fig.\ \ref{EB}, where both detectors exhibit identical efficiency $\eta_d$, and have the same level of electronic noise $\nu_{el}$. The electronic noise is modeled as a thermal state $\bar{n}_s$ with average photon number $\bar{n}_s$, where $\bar{n}_s=\nu_{el}/[2(1-\eta_{d})]$, and the efficiency is modeled as a beam splitter with transmittance $\eta_d$. 
\begin{figure}[t]
	\centering
	\includegraphics[width=7.5cm]{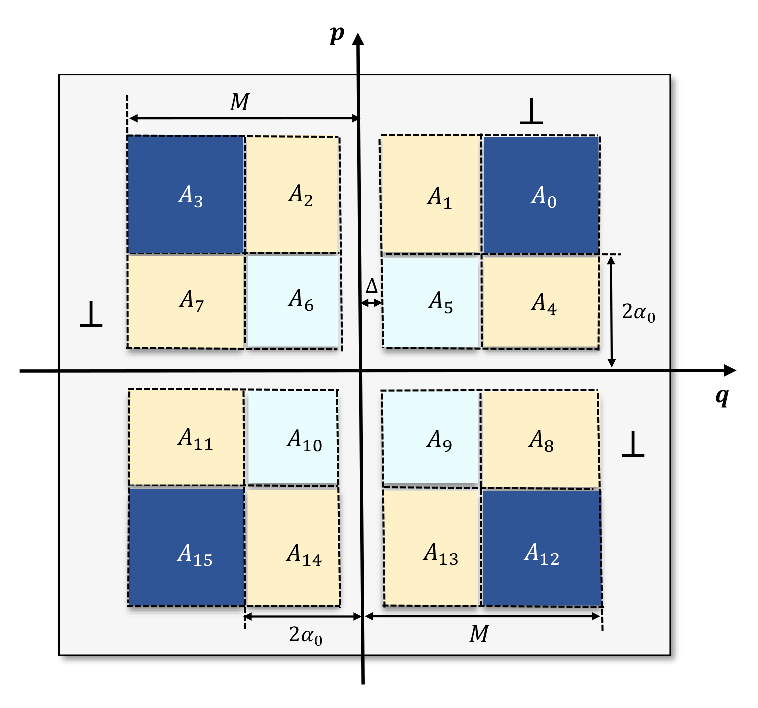}
	\caption{\label{xzt25}
		Bob's key map process involves assigning values to the measurement results $Y$. Each region $A_{z}$ corresponds to a specific key map value $z$, and $\alpha_0$ represents the distance between adjacent average states at Bob's side. During the post-selection phase, measurement results falling within a range of less than $\Delta$ from the coordinate axis or exceeding the detection limit are disregarded and instead marked with the symbol $\perp$.
	}
\end{figure}
The conditional entropy $H(X|E^{\prime})_{\bar{\rho}}$ of 16QAM-modulated protocol is detailed in the Appendix\ \ref{set11}. On this basis, the SDP can be described as
\begin{equation}\label{opt}
	\begin{aligned} 
		& \operatorname*{minimize}_{\bar{\rho}\in S^{\mathrm{E\&A}}}\ \
		H(X|E^{\prime})_{\bar{\rho}}\\
		&\text{subject to} \\
		&\hspace{0.1em}
		\begin{cases}
			\vspace{2pt}
			\hspace{0.5em}1-w \leq \mathrm{Tr}(\overline{\rho}) \leq 1,\\
			\vspace{2pt}
			\hspace{0.5em}	\frac{1}{2}\|\mathrm{Tr}_{\mathrm{B}}(\bar{\rho})-\tau_{A}\|_{1}\leq\sqrt{2w-w^{2}},\\
			\vspace{2pt}
			\hspace{0.5em}\mathrm{Tr}\left[\left(\frac{1}{P_{k}} |k\rangle\langle k|\otimes\hat{n}_{\beta_k}\right) \bar{\rho}\right] \leq \langle \hat{n}_{\beta_k} \rangle + \mu_{\hat{n}_{\beta_k}},\\
			\vspace{2pt}
			\hspace{0.5em}\mathrm{Tr}\left[\left(\frac{1}{P_{k}}|k\rangle\langle k|\otimes\hat{n}_{\beta_k}\right) \bar{\rho}\right] \geq \langle \hat{n}_{\beta_k} \rangle - \mu_{\hat{n}_{\beta_k}} - w\|\hat{n}_{\beta_k}\|_{\infty}, \\
			\vspace{2pt}
			\hspace{0.5em}\mathrm{Tr}\left[\left(\frac{1}{P_{k}}|k\rangle\langle k|\otimes\hat{n}_{\beta_k}^{2}\right) \bar{\rho}\right] \leq  \langle \hat{n}_{\beta_{k}}^{2} \rangle+\mu_{\hat{n}_{\beta_k}^{2}},  \\ 
			\vspace{2pt}
			\hspace{0.5em}\mathrm{Tr}\left[\left(\frac{1}{P_{k}}|k\rangle\langle k|\otimes \hat{n}^2_{\beta_k} \right) \bar{\rho}\right]\geq \langle \hat{n}_{\beta_k}^2 \rangle  - \mu_{\hat{n}^2_{\beta_k}} - w \|\hat{n}_{\beta_k}^2\|_{\infty},\\
			\vspace{2pt}
			\hspace{0.5em}\bar{\rho} \geq 0,
		\end{cases}
	\end{aligned}
\end{equation}
where $\hat{n}_{\beta_k}$ is the displaced photon number operator, $\hat{n}^{2}_{\beta_k}$ is the displaced squared photon number operator, $\beta_k=\sqrt{\eta}\alpha_{k}$ is the mean position of received state in the phase space after passing through the channel with transmittance $\eta$, 
the photon-number operator $\hat{n}=\hat{a}^{\dagger}\hat{a}$, and $X_\gamma=\hat{D}(\gamma)X\hat{D}^{\dagger}(\gamma)$, where $\hat{D}(\gamma)$ is the displacement operator with complex parameter $\gamma$. $\langle \hat{n}_{\beta_k} \rangle$ and $\langle \hat{n}_{\beta_k}^{2} \rangle$ are expectation of operators, which need to be calculated through measured result. Estimation method of the statistics is detailed in the Appendix\ \ref{set12}. $\mu_{\hat{n}_{\beta_k}}$ and $\mu_{\hat{n}^{2}_{\beta_k}}$ are parameters introduced by acceptance test which are defined as

\begin{equation}\label{opt}
	\mu_{\hat{n}_{\beta_k}}:={\sqrt{\frac{\|\hat{n}_{\beta_k}\|_{\infty}^{2}}{2k_{T}}\ln\left({\frac{2}{\epsilon_{\mathrm{AT}}}}\right)}},
\end{equation}
\begin{equation}\label{2}
	\mu_{\hat{n}^{2}_{\beta_k}}:={\sqrt{\frac{\|\hat{n}_{\beta_k}^2\|_{\infty}^{2}}{2k_{T}}\ln\left({\frac{2}{\epsilon_{\mathrm{AT}}}}\right)}}.
\end{equation}
$\tau_{A}$ is the quantum state of Alice's system which can be described as
\begin{equation}\label{3}
	\tau_{A}=\sum_{k,k^{\prime}=0}^{15}\sqrt{p_{k}p_{k^{\prime}}}\langle\varphi_{k^{\prime}}|\varphi_{k}\rangle|k\rangle\langle k^{\prime}|_{A}.
\end{equation}
In summary, the upper bound of composable secret key length against collective attacks can be obtained from Eqn.\ (\ref{eq:5}).

\subsection{Simulation method}
To evaluate the protocol performance, we simulate the quantum channel as a Gaussian channel with transmission $\eta_t$ and excess noise $\xi$, where $\eta_{t}=10^{-\alpha L/10}$ for transmission distance $L$ in kilometers, and $\alpha=0.2\ \text{dB/km}$ is the fiber loss. The excess noise $\xi$ is determined at the channel input, for example as preparation noise, so that Bob sees effective noise $\eta_{t}\xi$. 

According to the channel model, the statistical estimators used for SDP can be calculated as\ \cite{upadhyaya2021dimension}
\begin{equation}
	\langle \hat{n}_{\beta_k} \rangle = \frac{\eta_{t}\xi}{2},
\end{equation}
\begin{equation}
	\langle \hat{n}_{\beta_k}^{2} \rangle= \frac{\eta_{t}\xi(\eta_{t}\xi+1)}{2}.
\end{equation}
Cost of error correction is determined by the simulated joint probability distribution. When Alice prepares coherent state $\left|\alpha_{k}\right\rangle$, the probability of Bob obtaining key mapping result $z$ is given by the following integral:
\begin{equation}
	\begin{aligned}
		&P(Z=z|X=k)=\\
		&\int_{A_z} \frac{1}{\pi(1+\frac{1}{2}\eta_{d}\eta_{t}\xi+\nu_{\mathrm{el}})}\exp\left(\frac{-\left|y-\sqrt{\eta_{d}\eta_{t}}\alpha_{k}\right|^{2}}{1+\frac{1}{2}\eta_{d}\eta_{t}\xi+\nu_{\mathrm{el}}}\right) dy,
	\end{aligned}
\end{equation}
where $X$ and $Z$ represent Alice’s and Bob’s key strings, $k\in\left\{0,1,2,..., 15\right\}$, $z\in\left\{0,1,2,..., 15,\perp\right\}$, and $A_z$ is the post-selection range as shown in Appendix\ \ref{set11}.
Furthermore, error correction leakage can be bounded by
\begin{equation}
	\delta_{\mathrm{leak}}^{\mathrm{EC}} \leq p_{pass}\left\{n\left[(1-\beta)H(Z)+\beta H(Z|X) \right]+\log_{2}\left(\frac{2}{\epsilon_{EC}}\right)\right\}.
\end{equation}
Here, $n$ is the number of rounds used for key generation, $\beta$ is the reconciliation efficiency, $\log_{2}\left(2/{\epsilon_{EC}}\right)$ is the leaked information of correctness verification, and $p_{pass}$ is the probability that a round passes the post-selection.

Based on the above model, the numerical method in \ \cite{coles2016numerical,winick2018reliable} can be used to simulate the composable secret key rate of 16QAM-modulated CV-QKD protocol.

\begin{figure}[t]
	\centering
	\includegraphics[width=9cm]{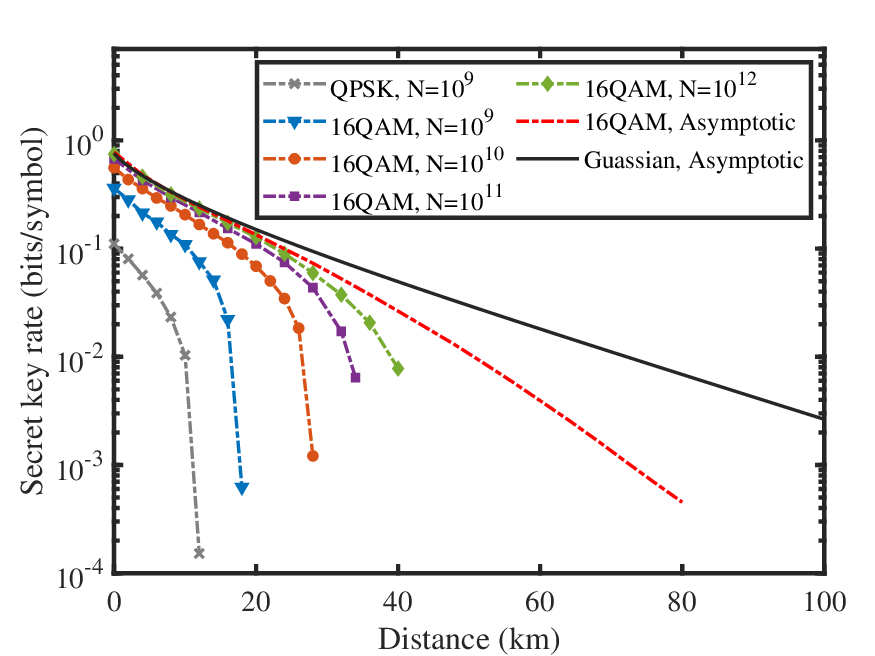}
	\caption{\label{16QAM_0}
		Composable secret key rate versus distance for QPSK-modulated CV-QKD, 16QAM-modulated CV-QKD, and Gaussian-modulated CV-QKD with ideal detectors. Different total number of signals $N$ are simulated and the testing ratios are fixed as $r_{test}= 10\%$. For QPSK-modulated protocol, modulation variance $V_A=0.49$ SNU. For 16QAM-modulated and Gaussian-modulated protocol, modulation variance $V_A=2$ SNU, excess noise $\xi = 0.01$, reconciliation efficiency $\beta=0.95$. The post-selection of discrete-modulated protocols is not considered ($\Delta=0$).
	}
\end{figure}

\begin{figure}[t]
	\centering
	\includegraphics[width=9cm]{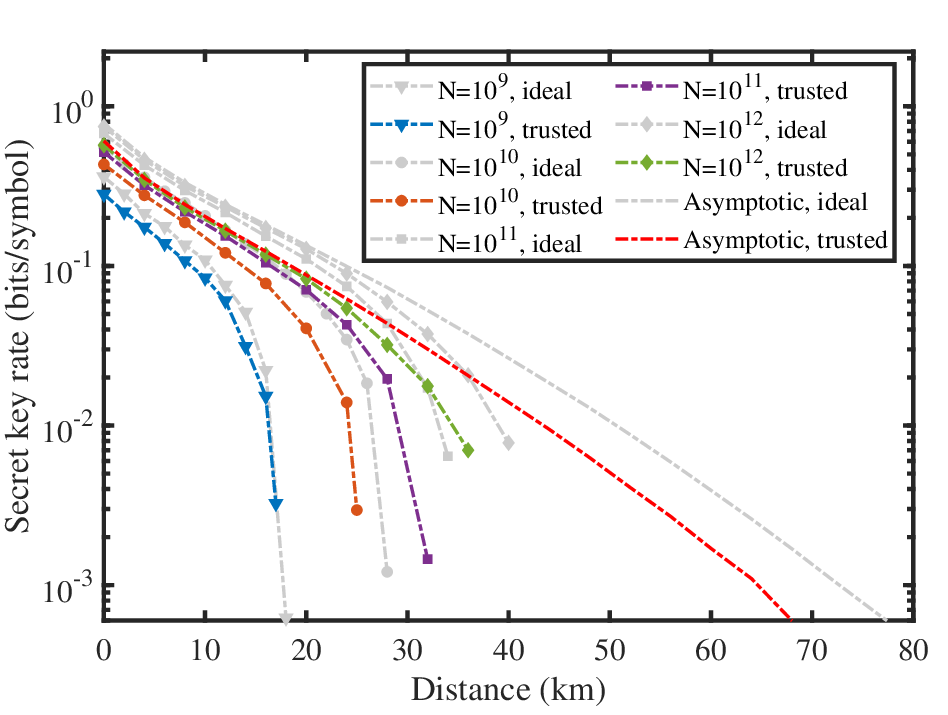}
	\caption{\label{16QAM_u}
		 Influence of detector model. Composable secret key rate versus distance for 16QAM-modulated
		CV-QKD with ideal detectors (gray dashed line) and trusted detectors (colored dashed line). Different total number of signals $N$ are simulated, and the testing ratios are fixed as $r_{test}= 10\%$. Modulation variance $V_A=2$ SNU, excess noise $\xi = 0.01$, detection efficiency $\eta_{d}=0.7$, detector noise $\nu_{el}=0.08$, reconciliation efficiency $\beta=0.95$, and post-selection is not considered ($\Delta=0$).
	}
\end{figure}
\subsection{Simulation results}
In the following simulation, we set the parameters to be $\epsilon_{EC}=2\times10^{-11}$, $\epsilon_{PA}=2\times10^{-11}$, $\epsilon_{AT}=7\times10^{-11}$, $\epsilon_{ET}=1\times10^{-11}$, $\bar{\epsilon}=1\times10^{-11}$, and the subspace dimension $N_c=10$ for a relatively fast computing\ \cite{upadhyaya2021dimension}. In order to match with practical experiments, unit of modulation variance here is SNU, where 1 SNU corresponds to variance of $0.5$ NU defined in the QPSK-modulated protocols\ \cite{lin2019,kanitschar2023finite}.

In Fig.\ \ref{16QAM_0}, the composable secret key rate of QPSK-modulated, 16QAM-modulated and Gaussian-modulated CV-QKD with ideal detector is shown against transmission distance. Different total number of signals $N=10^{9},10^{10},10^{11},10^{12}$ and asymptotic situation are simulated. The testing ratios are fixed as $r_{test}= 10\%$.  For QPSK-modulated protocol, modulation variance $V_A$ is set to be 0.49 SNU, which is close to its optimal value\ \cite{kanitschar2023finite}. For 16QAM-modulated and Gaussian-modulated protocol\ \cite{pirandola2024improved}, modulation variance $V_A=2$ SNU, excess noise $\xi = 0.01$, and reconciliation efficiency $\beta=0.95$. Post-selection of discrete-modulated protocols is not considered ($\Delta=0$). Simulation results indicate that, in asymptotic cases,, when transmission distance is small, the composable secret key rate achieved by the 16QAM-modulated protocol is virtually indistinguishable from that of the Gaussian-modulated protocol, given the same modulation variance. In terms of key rate, ignoring post-selection, it surpasses the CV-QKD protocol utilizing QPSK modulation with optimal modulation variance by approximately one order of magnitude under the total number of signals. When the total number of signals $N=10^{9}$, 16QAM-modulated CV-QKD protocol can achieve maximum transmission distance $L=18$ km. When the total number of signals reaches $10^{12}$, the maximum transmission distance can be increased to over 40 km.

In Fig.\ \ref{16QAM_u}, the composable secret key rate with ideal detectors and trusted detectors is shown versus the transmission distance. For the trusted, nonideal detectors, we set detection efficiency $\eta_{d}=0.7$ and detector noise $\nu_{el}=0.08$, other parameters are the same as Fig.\ \ref{16QAM_0}. Results show that compared to the ideal detector situation, the composable secret key rates with trusted detector have slightly decreased, and the maximum transmission distances have also been decreased. Even without considering post selection, a transmission distance of 25 km can still be achieved at $N=10^{10}$. When the total number of signals reaches $10^{12}$, the maximum transmission distance is limited to 40 km. These results indicate that the 16QAM-modulated protocol can achieve high-rate practical key distribution under short distance.

\begin{figure}[t]
	\centering
	\includegraphics[width=9cm]{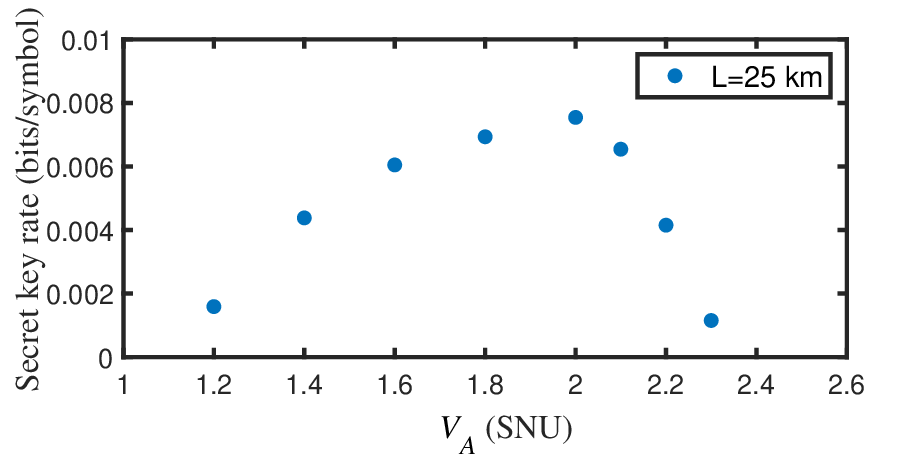}
	\caption{\label{VA}
		Composable secret key rate versus modulation variance $V_A$ for 16QAM-modulated CV-QKD with transmission distance $L=25$ km. The total number of signals $N=10^{10}$ and the testing ratio $r_{test}= 10\%$. Excess noise $\xi = 0.01$, detection efficiency $\eta_{d}=0.7$, detector noise $\nu_{el}=0.08$, reconciliation efficiency $\beta=0.95$, and post-selection is not considered.
	}
\end{figure}

\begin{figure}[t]
	\centering
	\includegraphics[width=9cm]{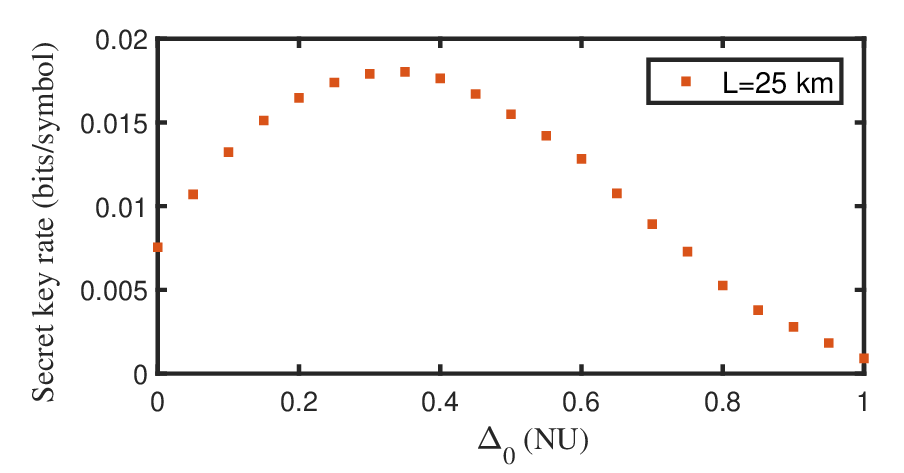}
	\caption{\label{DELTA}
		Optimization of post-selection parameter $\Delta_{0}$ for composable 16QAM-modulated CV-QKD with transmission distance $L=25$ km, respectively. The total number of signals $N=10^{10}$ and the testing ratio $r_{test}= 10\%$. Modulation variance $V_A=2$ SNU, excess noise $\xi = 0.01$, detection efficiency $\eta_{d}=0.7$, detector noise $\nu_{el}=0.08$, reconciliation efficiency $\beta=0.95$.
	}
\end{figure}

In Fig.\ \ref{VA}, modulation variance $V_A$ is optimized for 16QAM-modulated CV-QKD protocol with transmission distance $L=25$ km. Other parameters are the same as Fig.\ \ref{16QAM_u}. Results show that the optimal modulation variance is close to 2 SNU in the cases of $L=$ 25 km.

In Fig.\ \ref{DELTA}, post-selection parameter $\Delta_{0}$ is optimized for 16QAM-modulated CV-QKD protocol with transmission distance $L=25$ km, where $\Delta_{0}:=\Delta/\sqrt{\eta_{t}\eta_{d}}$. Modulation variance is set as $V_A=2$ SNU, and other parameters are the same as Fig.\ \ref{VA}. Results show that when the transmission distance is 25 km, the optimal post-selection parameter is around $\Delta_{0}=0.35$ NU, and compared to no post-selection ($\Delta_{0}=0$), the composable secret key rate can be increased by more than twice. This shows that a reasonable post-selection scheme can significantly improve the performance of 16QAM-modulated CV-QKD protocol.

\section{\label{set4}Experimental demonstration}
In this section, based on theoretical security analysis, experimental demonstration of the composable discrete-modulated CV-QKD protocol is provided. 

%
%
%

\begin{figure*}[t]
	\centering
	\includegraphics[width=16cm]{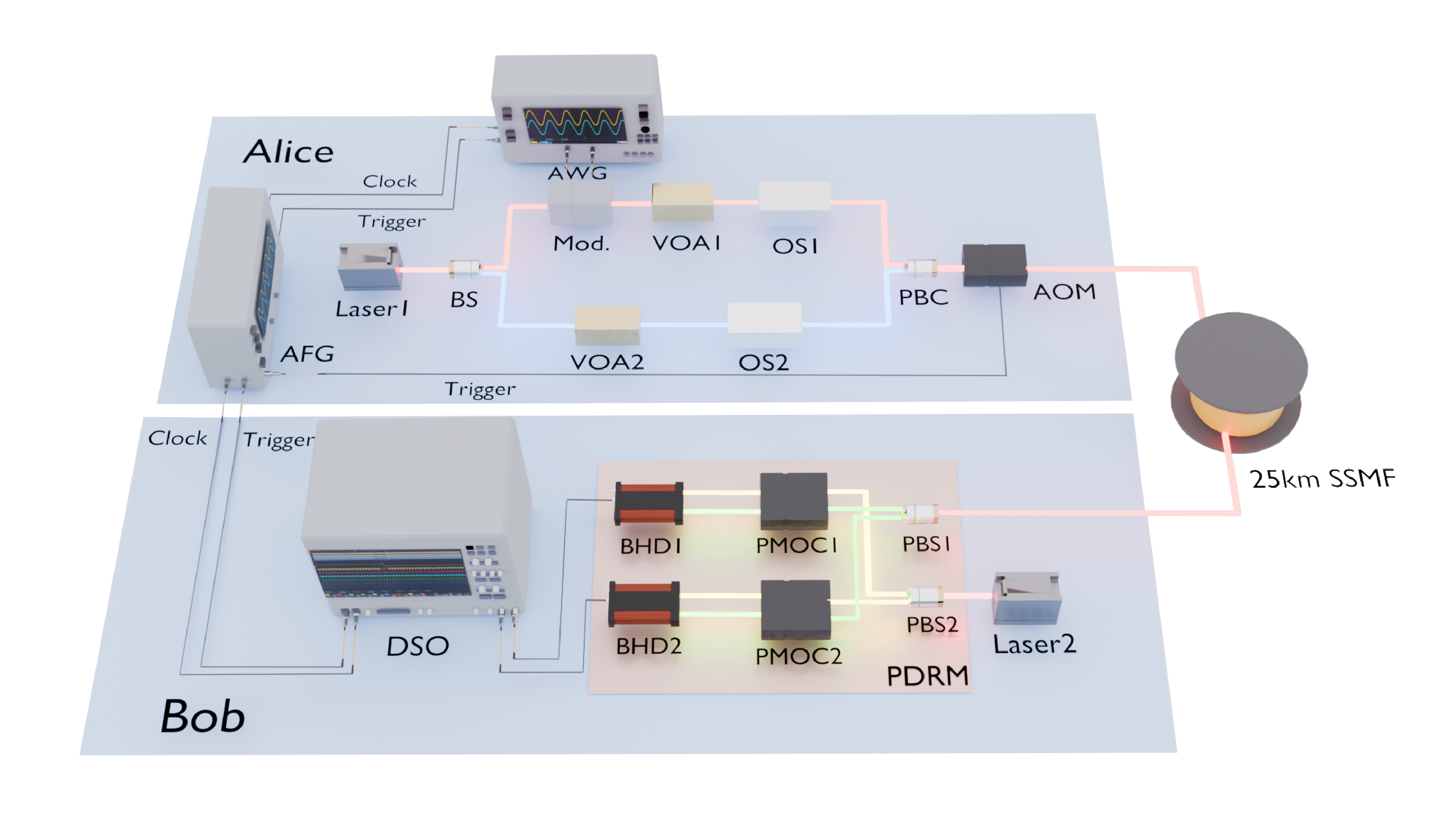}
	\caption{\label{e1}
		Experimental setup of the LLO discrete-modulated CV-QKD system. Mod., IQ modulator; BS, beam splitter; AWG, arbitrary waveform generator; VOA, variable optical attenuator; OS, optical switch; AFG, arbitrary function generator; PBC, polarization beam combiner; AOM, acousto-optic modulator; SSMF, standard single-mode fiber; PBS, polarization beam splitter; PMOC, polarization-maintaining optical coupler; BHD,
		balanced homodyne detector; PDRM, polarization diversity receiver module;  DSO, digital storage oscilloscope. 
	}
\end{figure*}

\subsection{Experimental setup}
The experimental setup for local local oscillator (LLO) discrete-modulated CV-QKD system is illustrated in Fig.\ \ref{e1}. On Alice's side, a continuous-wave laser emitting at 1550.22 nm with a linewidth below 100 Hz functions as the optical carrier. This laser beam is divided into two paths via a BS. One path traverses an In-phase/Quadrature modulator driven by a two-channel arbitrary waveform generator (AWG) operating at 10 GSa/s, featuring 10-bit resolution. This arrangement generates the quadrature components ($q$ and $p$) of probability-shaped 16QAM signals, with frequency shift of 1 GHz. These signals are modulated at a symbol rate of 1 Gbaud, employing a discrete Gaussian distribution and a root-raised cosine filter (roll-off factor of 0.3) for waveform shaped.

For channel training, QPSK training symbols, four times more powerful than the quantum signals, are interleaved with the quantum signals in the time domain. A variable optical attenuator (VOA1) adjusts the modulation variance, while an optical switch (OS1) manages the quantum link for parameter calibration, thereby generating discrete modulated coherent states. The reference path incorporates VOA2 and OS2, and both the discrete-modulated coherent states and reference signals are multiplexed in polarization and frequency. To minimize the impact of strong reference signal on the quantum signal, the reference signal's frequency is shifted to a region with weaker detector response, while maintaining sufficiently high signal-to-noise ratio (SNR). Furthermore, frequency separation between the quantum and reference signals can be increased.

The multiplexed signal then passes through an acoustic-optic modulator (AOM) for real-time shot noise calibration. The transmission path comprises 25 km of standard single-mode fiber (SSMF). On Bob's side, a second independent continuous-wave laser with a linewidth of $\textless$100 Hz serves as the local oscillator (LO), detuned by approximately 2 GHz from Alice's laser. The signal and LO are coherently detected using a polarization diversity receiver module (PDRM). It consists of two polarization beam splitters (PBS), two polarization-maintaining optical couplers (PMOC), and two balanced homodyne detectors (BHD), offering 3 dB bandwidth of 1.6 GHz, responsivity of 0.95 A/W, and gain of $3.0\times10^{4}$ V/A. Notably, this experimental setup eliminates the need for active polarization state control of the incoming light signal at the receiver. Instead, full DSP is leveraged for polarization compensation, enhancing system robustness and ensure that the CV-QKD system operates with minimal excess noise, thereby enhancing its performance and security.

Ultimately, electrical signals are digitized by a digital storage oscilloscope (DSO) operating at 8 GSa/s with 10-bit resolution, followed by offline digital signal processing (DSP) to recover and analyze the raw data. Here, clock signals for both AWG and DSO are provided by a 10 MHz sine wave generated by an arbitrary function generator (AFG). Simultaneously, the AFG outputs a 50$\%$ duty cycle pulse signal to synchronize the AWG, AOM, and DSO, enabling coordinated generation, control, and acquisition of signals.

The execution of DSP algorithms is intricate and detailed, encompassing the following steps: Firstly, the algorithm extracts the reference signal based on the stronger of the two varying intensity signals received along the principal axis of the PDRM, which are caused by random polarization fluctuations. Next, it estimates the frequency offset in the frequency domain. To suppress out-of-band noise, bandpass filtering with bandwidths of 1.3 GHz and 200 kHz is applied to the quantum and reference signals, respectively. Frequency shifts are compensated for by digitally shifting the reference signal by 1 GHz to align the center frequencies. Demodulation of the q and p components of quantum states is performed digitally from the intermediate frequency signal, while simultaneously compensating for carrier frequency shifts and phase noise introduced by Alice's and Bob's lasers using the extracted reference signal. The demodulated signal is then resampled to achieve 4 times oversampling, followed by the application of matched filtering with a root-raised cosine (RRC) filter. Lastly, $4\times2$ multiple input multiple output (MIMO) equalization is employed to compensate for polarization fluctuations and imbalances in the $q$ and $p$ components. This step also involves downsampling the quantum signal to one sample per symbol and reducing residual noise. Denoting $T S_{i n q}^{X}$, $T S_{inp}^{X}$, $T S_{i n q}^{Y}$, $T S_{i n p}^{Y}$ are the received training symbols, then, the training processing can be described by
\begin{equation}\label{opt}
	\begin{aligned}
		T S_{outq}^{X}=\omega_{11}T S_{i n q}^{X}+\omega_{12}T S_{i n p}^{X}+\omega_{13}T S_{i n q}^{Y}+\omega_{14}T S_{i n p}^{Y}, \\
		T S_{outp}^{X}=\omega_{21}T S_{i n q}^{X}+\omega_{22}T S_{i n p}^{X}+\omega_{33}T S_{i n q}^{Y}+\omega_{24}T S_{i n p}^{Y}.
	\end{aligned}
\end{equation}
the update processing of the tap coefficients $\omega_{1j}(j\in\{1,2,3,4\})$ is 
\begin{equation}\label{opt}
	\begin{aligned}
		\omega_{1j}(m+1)=\omega_{1j}(m)+\mu\varepsilon_{q}^{X}(n)T S_{outq}^{X}, \\
		\omega_{2j}(n+1)=\omega_{2j}(n)+\mu\varepsilon_{p}^{X}(n)T S_{outp}^{X}.
	\end{aligned}
\end{equation}
where $\mu$ is the step-size, $\varepsilon_{q}^{X}(n)$ and $\varepsilon_{p}^{X}(n)$ are the error value, which can be calculated by
\begin{equation}\label{opt}
	\begin{aligned}
		\varepsilon_{q}^{X}=TS_{q}^{X}-TS_{out q}^{X}, \\
		\varepsilon_{p}^{X}=TS_{p}^{X}-TS_{out p}^{X}.
	\end{aligned}
\end{equation}
where $TS_{q}^{X}$ and $TS_{p}^{X}$ are the transmitted training symbols. Notably, to improve the convergence speed and accuracy of the LMS algorithm, SNR enhancement of the training sequence is performed by time-domain superposition method\ \cite{pan2024high2}. Finally, the raw data is achieved to evaluate the system performance.

\begin{table}[t]
	\centering
	\caption{Experimental parameters and estimated results.}
	\renewcommand{\arraystretch}{1.4}
	\resizebox{1\hsize}{!}{
		\begin{tabular}{|c|c|c|}
			\hline
			Parameter                    & Symbol    & Value                                   \\ \hline
			The total data amount & $N$            & $1.28\times10^{10}$                                            \\ \hline
			The test data amount & $m$            & $6.4\times10^{9}$                                       \\ \hline
			Modulation variance & $V_{A}$            & $2.03$ (SNU)                                            \\ \hline
			Probability distribution parameter & $\nu$ & $0.2$                                               \\ \hline
			Worst-case channel transmittance          & $T_{wc}$                 & $0.3465$                                           \\ \hline
			Worst-case excess noise           & $\xi_{wc}$ & $0.0083$                                              \\ \hline
			Electronic noise           & $\nu_{\mathrm{el}}$ & 0.0883                                           \\ \hline
			Detection efficiency    & $\eta_d$   & 0.714                                          \\ \hline
			Displaced photon number                   & $\langle \hat{n}_{\beta_k} \rangle$                 & 0.0012   \\ \hline
			Displaced squared photon number                   & $\langle \hat{n}_{\beta_k}^2 \rangle$                 &  0.0012 \\ \hline
			Transmission distance                   & $L$                 & 25 (km) \\ \hline
			Fiber loss                   & $\alpha$                 & 0.184 (dB/km) \\ \hline
			Post-selection parameter                   & $\Delta_{0}$                 & $0.35$ (NU) \\ \hline
			Reconciliation efficiency                   & $\beta$                 & $0.95$  \\ \hline
			Frame error rate                   & FER                 & $0.15$  \\ \hline
			Training sequences ratio                   & $a$                 & $0.25$  \\ \hline
			System repetition frequency                   & $R$                 & $1$ (GHz)  \\ \hline
			Total security parameter                   & $\epsilon$                 & $10^{-9}$  \\ \hline
		\end{tabular}
	}
	\label{tab1}
\end{table}

\subsection{Experimental results}
We collect 20 sets of data blocks, and use 10 sets for test, each block contains 640 M symbols. 
In order to evaluate the performance of our system, we consider two situations: the secret key rate with Gaussian channel assumption and the secret key rate in general situation. It is worth mentioning that Gaussian attack is not optimal in discrete-modulated CV-QKD protocols. However, in our laboratory system, there is no actual eavesdroppers, and we can still consider the channel as Gaussian to estimate channel transmittance and excess noise to evaluate system performance preliminarily attributing to theoretical results. In practical situations, the system may be attacked by eavesdroppers, which may be non-Gaussian and results in a worse key rate. Therefore, it is necessary to estimate the statistical estimators $\langle \hat{n}_{\beta_k} \rangle$ and $\langle \hat{n}_{\beta_k}^{2} \rangle$ required for SDP using experimental data directly, and obtain a more reliable secret key rate.

\begin{figure}[t]
	\centering
	\includegraphics[width=9cm]{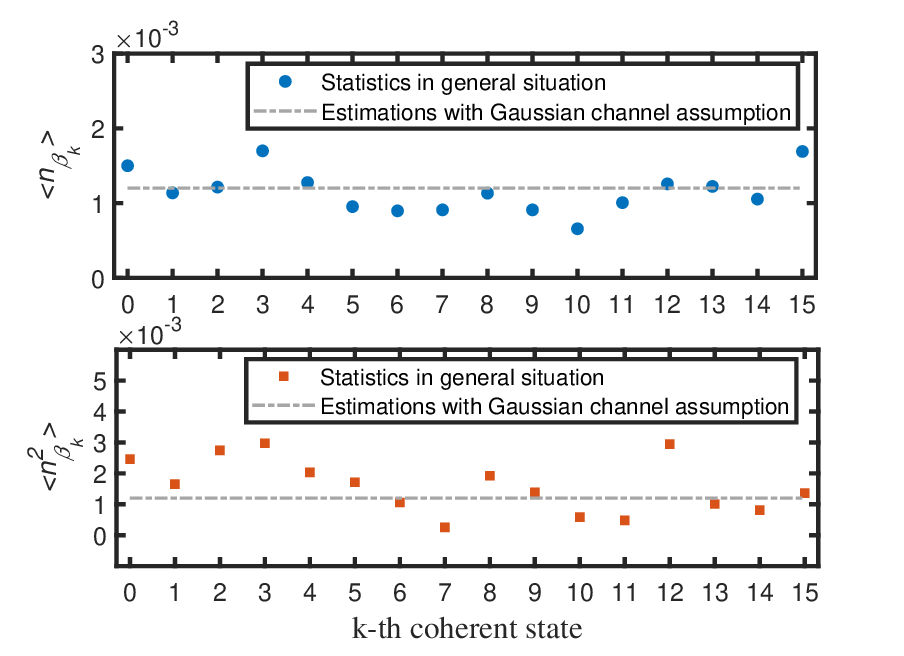}
	\caption{\label{n}
		Experimental estimation of the statistics $\langle \hat{n}_{\beta_k} \rangle$ and $\langle \hat{n}_{\beta_k}^{2} \rangle$. The dots are the statistics in general situation, and the line is the estimations with Guassian channel assumption.
	}
\end{figure}

The statistical estimators $\langle \hat{n}_{\beta_k} \rangle$ and $\langle \hat{n}_{\beta_k}^{2} \rangle$ are estimated in general situation and compared to the value with Gaussian channel assumption as shown in Fig.\ \ref{n}.
The statistics $\langle \hat{n}_{\beta_k} \rangle$ and $\langle \hat{n}_{\beta_k}^{2} \rangle$ fluctuate around the estimated values  with Gaussian channel assumption due to measurement and statistical errors, and the fluctuation of $\langle \hat{n}_{\beta_k}^{2} \rangle$ is greater. Since there is no Eve in our laboratory, when the amount of data is infinite, the statistical values in general situation should be infinitely close to the experimental estimations with Gaussian channel assumption.

The experimental parameters and estimated results are summarized in Tab.\ \ref{tab1}. With the security parameters $\epsilon_{EC}=2\times10^{-11}$, $\epsilon_{PA}=2\times10^{-11}$, $\epsilon_{AT}=7\times10^{-11}$, $\epsilon_{ET}=1\times10^{-11}$ and $\bar{\epsilon}=1\times10^{-11}$, the security parameter of each block is $\epsilon=10^{-10}$, and the total security parameter is $\epsilon=10^{-9}$. In order to optimize the performance of our system, we optimize the post-selection parameter $\Delta_{0}$ using the parameters in Gaussian channel and decide the optimal post-selection parameter is around 0.35 NU.

\begin{figure}[t]
	\centering
	\includegraphics[width=8.5cm]{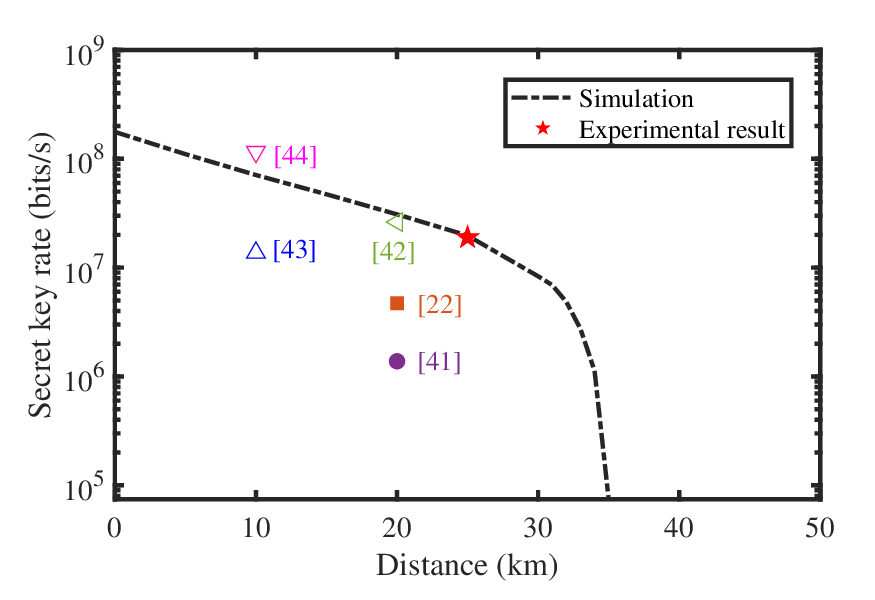}
	\caption{\label{skr}
		Composable secret key rate versus transmission distance with experimental parameters in Table\ \ref{tab1}. The red star is general experimental secret key rate, the black curve is the simulation using experimental parameters with Guassian channel assumption. The solid orange square represents the experimental result of previous Guassian-modulated CV-QKD systems corresponding Ref.\ \cite{jain2022practical}. The solid purple circle represents the QPSK-modulated CV-QKD system  during the same period corresponding Ref.\ \cite{hajomer2024experimental}. The hollow triangles with square bracket numbers represent the experimental results of previous discrete-variable QKD systems corresponding Ref.\ \cite{islam2017provably,yuan201810,li2023high}. 
	}
\end{figure}

Based on the optimized post-selection parameter, we calculate the secret key rates under Gaussian channel assumption and general situation, respectively, as shown in Fig.\ \ref{skr}. Considering reconciliation efficiency $\beta=0.95$, frame error rate FER=0.15, system repetition frequency $R=1$ GHz,  and $25 \% $ of the keys are used for training sequences, the experimental key rate in general situation is 18.93 Mbps as shown by the red star. The black curve is the simulation result using experimental parameters with Gaussian channel assumption as a reference for system performance. Compared to previous Guassian-modulated CV-QKD systems featuring composable security, the 16QAM-modulated system demonstrates improvement of more than one order of magnitude in terms of secret key rate and achieves longer transmission distance\ \cite{jain2022practical}. Furthermore, when compared to the CV-QKD system employing QPSK modulation, our system exhibits a performance advantage of nearly two orders of magnitude\ \cite{hajomer2024experimental}. Compared to previous high-rate discrete-variable QKD systems with composable security\ \cite{islam2017provably,yuan201810,li2023high}, our system can also achieve a comparable level of secret key rate. Results indicate that the CV-QKD system  with 16QAM modulation can ensure high composable key rates over medium to short distances while maintaining low complexity and cost, further demonstrating the practical advantages in urban areas.

\section{\label{set5}Discussion and conclusion}
Over recent years, the theoretical security of discrete-modulated CV-QKD has been progressively refined. The method of nonlinear SDP makes the constraint of secret key rate become more compact, and can use advanced tools such as entropy accumulation theorem\ \cite{dupuis2020entropy} to achieve security proof under coherent attacks. However, a significant limitation of these protocols is that as the number of modulation constellations increases, so does the matrix dimension required for SDP solution, leading to a substantial rise in computational time. It is difficult to extend to higher-order modulation formats such as 64QAM and above with existing computing resources. Therefore, there is an urgent need to improve algorithms for solving optimization problems in order to reduce computational complexity. Although some improved algorithms have been proposed\ \cite{hu2022robust,karimi2023efficient,kossmann2024optimising}, the effectiveness still falls short of meeting the requirements of high-order modulation.

Furthermore, the protocol utilizing nonlinear SDP method necessitates Bob to discretize his data. However, this presents a challenge as the exact values of his measurement results cannot be utilized for information reconciliation. Consequently, traditional Gaussian modulation coordination schemes become inapplicable in such protocols, leaving the question of how to achieve efficient information reconciliation\ \cite{leverrier2023information}. Moreover, this makes the method of using all data for parameter estimation and key extraction in Guassian-modulated protocol\ \cite{jain2022practical} no longer applicable. As a result, significant key consumption arises with finite-size effect. These are the theoretical challenges that need to be solved for the practical application of discrete-modulated CV-QKD protocols, which are left for future work.

In conclusion, in order to address the issue of low secret key rate in CV-QKD systems with composable security, we propose the probability-shaped 16QAM-modulated CV-QKD protocol. Theoretical performance is analyzed and experimental demonstration is provided. Performance of 16QAM-modulated protocol significantly surpasses that of QPSK modulation, and is comparable to Gaussian-modulated protocol at close range. Specifically, our system achieves composable secret key rate of 18.93 Mbps over 25 km fiber channel. The secret key rate exceeds that of previous Gaussian-modulated CV-QKD systems by more than an order of magnitude, exceeds the QPSK CV-QKD system of the same period by nearly two orders of magnitude, and is comparable to high-rate discrete-variable QKD systems, all while maintaining low system complexity and cost. Our work offers a valuable solution for the future deployment of QKD.

\begin{acknowledgments}
	The authors thank Florian Kanitschar for valuable discussions about theoretical security analysis and numerical calculation in the early preparation stage. We acknowledge financial support from the National Key Research and Development Program of China (Grant No. 2020YFA0309704), the National Natural Science Foundation of China (Grants No. U24B2013, U22A2089, 62471446, 62301517, 62101516, 62171418, 62201530, 62001044), the Sichuan Science and Technology Program (Grants No. 2024ZYD0008, 2024JDDQ0008, 2023ZYD0131, 2023JDRC0017, 2022ZDZX0009, 2023NSFSC1387, 2024NSFSC0470, and 2024NSFSC0454), the National Key Laboratory of Security Communication Foundation(Grant No. 6142103042301, 6142103042406), Stability Program of National Key Laboratory of Security Communication(Grant No. WD202413, WD202414), the Basic Research Program of China (Grant No. JCKY2021210B059), the Equipment Advance Research Field Foundation (Grant No. 315067206).
\end{acknowledgments}

\appendix

\section{\label{set11}Region operators of 16QAM-modulated CV-QKD}
In the security analysis framework\ \cite{coles2016numerical,winick2018reliable}, 
the conditional entropy $H(X|E^{\prime})_{\bar{\rho}}$ is usually represented as $D\big(\mathcal{G}\big(\bar{\rho}\big)||\mathcal{Z}\big[\mathcal
{G}\big(\bar{\rho}\big)\big]\big)$, where $\mathcal{G}$ is a completely positive and trace-preserving (CPTP) map that outlines several classical post-processing procedures associated with the protocol, $\mathcal{Z}$ is a pinching quantum channel, $D(\rho||\sigma)$  is the quantum relative entropy. In the case of reverse reconciliation, ${\cal G}(\sigma)=K\sigma K^{\dagger}$, where $K=\sum_{z=0}^{15}|z\rangle_{R}\otimes I_{A}\otimes(\sqrt{R^{z}})_{B}$ and $R^{z}$ are region operators. The region operators are defined as ${R}_{B}^{z}=\frac{1}{\pi}\int_{A^{z}} |\zeta\rangle \langle \zeta| d^{2}\zeta$, where $A^{z}$ are the regions in phase space.
For the trusted noise scenario, the noisy region operators are\ \cite{lin2020trusted}
\begin{equation}
[{R}_{B}^{z}]'=\textstyle\int_{\zeta\in A^{z}}G_{\zeta}d^{2}\zeta,
\end{equation}
and the key map POVM elements
\begin{equation}
	\left[{P}^{z}\right]^{\prime}=I_{A}\otimes\left[{R}_{B}^{z}\right]^{\prime}.
\end{equation}
Because the base of our protocol is $\{|i\rangle_{A}\otimes|n_{\beta_{i}}\rangle_{B}\}$, the matrix elements of the POVM are\ \cite{upadhyaya2021dimension}
\begin{equation}
	\begin{aligned}
		&\left[{P}^{z}\right]_{k l m n}^{\prime}=\langle m_{\beta_{l}}|\left[{ R}_{B}^{z}\right]^{\prime}|n_{\beta_{k}}\rangle\,,\\
		&=\int_{A^{z}}\langle m_{\beta_{l}}|\,G_{\zeta}\,|n_{\beta_{k}}\rangle\,d^{2}\zeta,\\
		&={\frac{1}{\eta_{d}\pi}}\int_{A^{z}}\left\langle m\right|D\left({\frac{\zeta}{\sqrt{\eta_{d}}}}-\beta_{k}\right)\rho_{t h}({\bar{n}})D^{\dagger}\left({\frac{\zeta}{\sqrt{\eta_{d}}}}-\beta_{k}\right) \left|n\right\rangle d^{2}\zeta,\\
		&=\int_{A^{z}}\,\langle m|\,G_{\zeta-\sqrt{\eta_{d}}\beta_{k}}\,|n\rangle\,d^{2}\zeta,
	\end{aligned}
\end{equation}
The integral function of POVM element $G_{\zeta-\sqrt{\eta_{d}}\beta_{k}}$ in photon-number basis is
\begin{equation}
	\begin{aligned}
		&\left\langle m\right|G_{\zeta-\sqrt{\eta_{d}}\beta_{k}}\left|n\right\rangle      =\frac{1}{\eta_{d}\pi}\exp\left[\frac{-|\zeta-\sqrt{\eta_{d}}\beta_{k}|^{2}}{\eta_{d}(1+\bar{n}_{d})}\right]\frac{\bar{n}_{d}^{m}}{(1+\bar{n}_{d})^{n+1}} \\
		&\left(\frac{\left(\zeta-\sqrt{\eta_{d}}\beta_{k}\right)^{*}}{\sqrt{\eta_{d}}}\right)^{n-m}\left(\frac{m!}{n!}\right)^\frac{1}{2}L_{m}^{(n-m)}\left[-\frac{\vert \zeta-\sqrt{\eta_{d}}\beta_{k}\vert^{2}}{\eta_{d}\bar{n}_{d}(1+\bar{n}_{d})}\right],
	\end{aligned}
\end{equation}
where ${\bar{n}}_{d}=(1-\eta_{d}+\nu_{\mathrm{el}})/\eta_{d}$, $L_{a}^{(b)}(c)$ is the generalized Laguerre polynomial of degree $a$ with a parameter $b$ in the variable $c$.

For our 16QAM-modulated protocol, we use cartesian coordinate system, and $\zeta=x+iy$. The integration limits $A_z$ are 
\begin{equation}
	\int_{\Delta_{y,low}}^{\Delta_{y,up}}\int_{\Delta_{x,low}}^{\Delta_{x,up}}\langle m|\,G_{x+iy-\sqrt{\eta_{d}}\beta_{k}}\,|n\rangle dxdy,
\end{equation}
where 
\begin{equation}
	\begin{aligned}
		\Delta_{xlow}=&\left\{2\alpha_{0},\Delta,-2\alpha_{0},-\infty,2\alpha_{0},\Delta,-2\alpha_{0},-\infty, \right.\\
		&\left.2\alpha_{0},\Delta,-2\alpha_{0},-\infty,2\alpha_{0},\Delta,-2\alpha_{0},-\infty \right\}, \\
		\Delta_{xup}=&\left\{\infty,2\alpha_{0},-\Delta,-2\alpha_{0},\infty,2\alpha_{0},-\Delta,-2\alpha_{0},\right.\\
		&\left.\infty,2\alpha_{0},-\Delta,-2\alpha_{0},\infty,2\alpha_{0},-\Delta,-2\alpha_{0} \right\},\\
		\Delta_{ylow}=&\left\{2\alpha_{0},2\alpha_{0},2\alpha_{0},2\alpha_{0},\Delta,\Delta,\Delta,\Delta,\right.\\
		&\left.-2\alpha_{0},-2\alpha_{0},-2\alpha_{0},-2\alpha_{0},-\infty,-\infty,-\infty,-\infty \right\},\\
		\Delta_{yup}=&\left\{\infty,\infty,\infty,\infty,2\alpha_{0},2\alpha_{0},2\alpha_{0},2\alpha_{0},\right.\\
		&\left.-\Delta,-\Delta,-\Delta,-\Delta,-2\alpha_{0},-2\alpha_{0},-2\alpha_{0},-2\alpha_{0}\right\}.
	\end{aligned}
\end{equation}

\section{\label{set12}Statistical estimation}

Firstly, we use traditional parameter estimation methods to estimate channel transmittance $T$ and excess noise $\xi$ of the system. We model the quantum channel as an additive Gaussian white noise channel, for heterodyne detection scheme, that satisfies
\begin{equation}
	y={\sqrt{0.5\eta_{d}T}}x+\delta,
\end{equation}
where $x$ and $y$ represent the input and output of the channel, $\delta$ is the Gaussian noise with variance $T\eta_d\xi/2+1+\nu_{el}$, $T$ is the channel transmittance, and $\eta_d$ is the detection efficiency. In this case, the channel transmittance $T$ and excess noise $\xi$ can be estimated as
\begin{equation}
	T={\frac{\left({\sum_{i=1}^{m}x_{i}y_{i}/m}\right)^{2}}{0.5\eta_{d}}},
\end{equation}
\begin{equation}
	\xi={\frac{V_{B}-0.5\eta_{d}T V_{A}-\nu_{e l}-1}{0.5\eta_{d}T}},
\end{equation}
where $V_{A}$ represents the modulation variance of Alice, $V_{B}$ represents the variance of Bob's data, $\nu_{el}$ is the variance of detector electrical noise, and $m$ is the amount of data used for tests. Due to the fluctuation of statistical values under finite-size effect, we consider the worst-case estimate\ \cite{pirandola2024improved}, 
\begin{equation}
	T_{\mathrm{wc}}\simeq T\ -w\frac{2T}{\sqrt{2k_{T}}}\sqrt{\frac{\xi+\frac{2+\nu_{e l}}{\eta_{d}T}}{V_{A}}},
\end{equation}
\begin{equation}
	\xi_{\mathrm{wc}}\simeq{\frac{T}{T_{\mathrm{wc}}}}\,\xi+w\sqrt{\frac{1}{k_{T}}}\frac{\eta_{d}T\xi+2+\nu_{e l}}{\eta_{d}T_{\mathrm{wc}}},
\end{equation}
the parameter $w$ can be given by the inverse error function and is related to the security parameter $\epsilon_{pe}$ as
\begin{equation}
	w=\sqrt{2}\mathrm{erf}^{-1}(1-\epsilon_{pe}).
\end{equation}

Due to the fact that Gaussian attacks are not optimal for discrete-modulated CV-QKD protocols, general statistical estimations are also necessary. In order to calculate the statistics $\langle \hat{n}_{\beta_k} \rangle$ and $\langle \hat{n}_{\beta_k}^{2} \rangle$, we first distinguish the number of rounds for sending each state $\left|\alpha_{k}\right\rangle$, denoted as $C_{k}$. The average measurement value of each state corresponding to Bob is
\begin{equation}
	\bar{Y}^k=\frac{1}{C_{k}}\sum_{j=1}^{C_{k}}\left(q_{j}^{k}+i p_{j}^{k}\right).
\end{equation}
We can calculate the displaced values 
\begin{equation}
	\tilde{q}_{j}^{k}=\,q_{j}^{k}\,-\mathrm{Re}(\bar{Y}^{k}),
\end{equation}
\begin{equation}
	\tilde{p}_{j}^{k}=p_{j}^{k}-\mathrm{Im}(\bar{Y}^{k}).
\end{equation}
Because the expected value of displaced observation calculated on original undisplaced data is the same as the expected value of undisplaced observation calculated on the displaced data, according to further deductions, we can calculate the statistics with detector noise\ \cite{upadhyaya2021tools}
\begin{equation}
	\langle[\hat{n}_{\sqrt{\eta_{d}}\beta_{k}}]^{\prime}\rangle=\frac{1}{C_{k}}\sum_{j=1}^{C_{k}}\left[\frac{1}{2}(\tilde{q}_{j}^{k})^{2}+\frac{1}{2}(\tilde{p}_{j}^{k})^{2}-1\right],
\end{equation}
\begin{equation}
	\begin{aligned}
		\left\langle[\hat{n}_{\sqrt{\eta_{d}}\beta_{k}}^{2}]^{\prime}\right\rangle=&\frac{1}{C_{k}}\sum_{j=1}^{C_{k}}\Bigl[\frac{1}{4}(\tilde{q}_{j}^{k})^{4}+\frac{1}{2}(\tilde{q}_{j}^{k})^{2}(\tilde{p}_{j}^{k})^{2}+\\
		&\frac{1}{4}(\tilde{p}_{j}^{k})^{4}-\frac{3}{2}(\tilde{q}_{j}^{k})^{2}-\frac{3}{2}(\tilde{p}_{j}^{k})^{2}+1\Bigr].
	\end{aligned}
\end{equation}
According to this, we can reconstruct the effective ideal expectations as\ \cite{upadhyaya2021dimension}
\begin{equation}
	\langle{\hat{n}}_{\beta_{k}}\rangle=\frac{\langle[{\hat{n}}_{\sqrt{\eta_{d}}\beta_{k}}]^{\prime}\rangle-\nu_{e l}}{\eta_{d}},
\end{equation}
\begin{equation}
	\begin{aligned}
		\langle{\hat{n}}_{\beta_{k}}^{2}\rangle=&\frac{1}{\eta_{d}^{2}}\biggl(\langle[{\hat{n}}_{\sqrt{\eta_{d}}\beta_{k}}^{2}]^{'}-2\nu_{e l}^{2}-\nu_{e l}-\\
		&(4\nu_{el}+1-\eta_{d})\left(\left\langle[\hat{n}_{\sqrt{\eta_{d}}\beta_{k}}^{2}]^{\prime}\right\rangle-\nu_{e l}\right)\biggr).
	\end{aligned}
\end{equation}
These statistics can be incorporated into SDP to calculate the secret key rate.


\bibliography{ref}

\end{document}